\documentclass[11pt]{article}
\usepackage{moriond,epsfig}
\usepackage{subfigure}
\usepackage{xspace}

\def\Journal#1#2#3#4{{#1} {\bf #2}, #3 (#4)}

\def\PLB{{\em Phys. Lett.}  B}
\def\PRL{\em Phys. Rev. Lett.}
\def\PRD{{\em Phys. Rev.} D}
\def\EPJC{{\em Eur. Phys. J.} C}
\def\JHEP{{\em J. High Energy Phys.}}

\def\be{\begin{equation}}
\def\ee{\end{equation}}
\def\bea{\begin{eqnarray}}
\def\eea{\end{eqnarray}}
\newcommand{\dzero} {D0\xspace}
\newcommand{\MET}{\ensuremath{\not \!\! E_T}}
\begin{document}
\begin{flushright}
FERMILAB-CONF-12-201-PPD
\end{flushright}

\vspace*{3.8cm}
\title{SINGLE TOP PRODUCTION AT THE TEVATRON}

\author{ ZHENBIN WU \\ (on behalf of the CDF and \dzero collaborations) }

\address{Department of Physics, Baylor University, \\ One Bear Place \#97316,
Waco TX 76798-7316, USA}

\maketitle\abstracts{
We present recent results of single top quark production in the lepton plus
jet final state, performed by the CDF and \dzero collaborations based on 7.5
and 5.4~fb$^{-1}$ of $p\bar p$ collision data collected at $\sqrt{s} = 1.96$
TeV from the Fermilab Tevatron collider. Multivariate techniques are used to
separate the single top signal from the backgrounds. Both collaborations
present measurements of the single top quark cross section and the CKM matrix
element $|V_{tb}|$. A search for anomalous $Wtb$ coupling from \dzero is also
presented.}

\section{Introduction}
\begin{figure}[ht!]
  \centering
  \subfigure[\label{fig:t-channel_NLO} $t$-channel]{
  \includegraphics[width=0.3\textwidth]{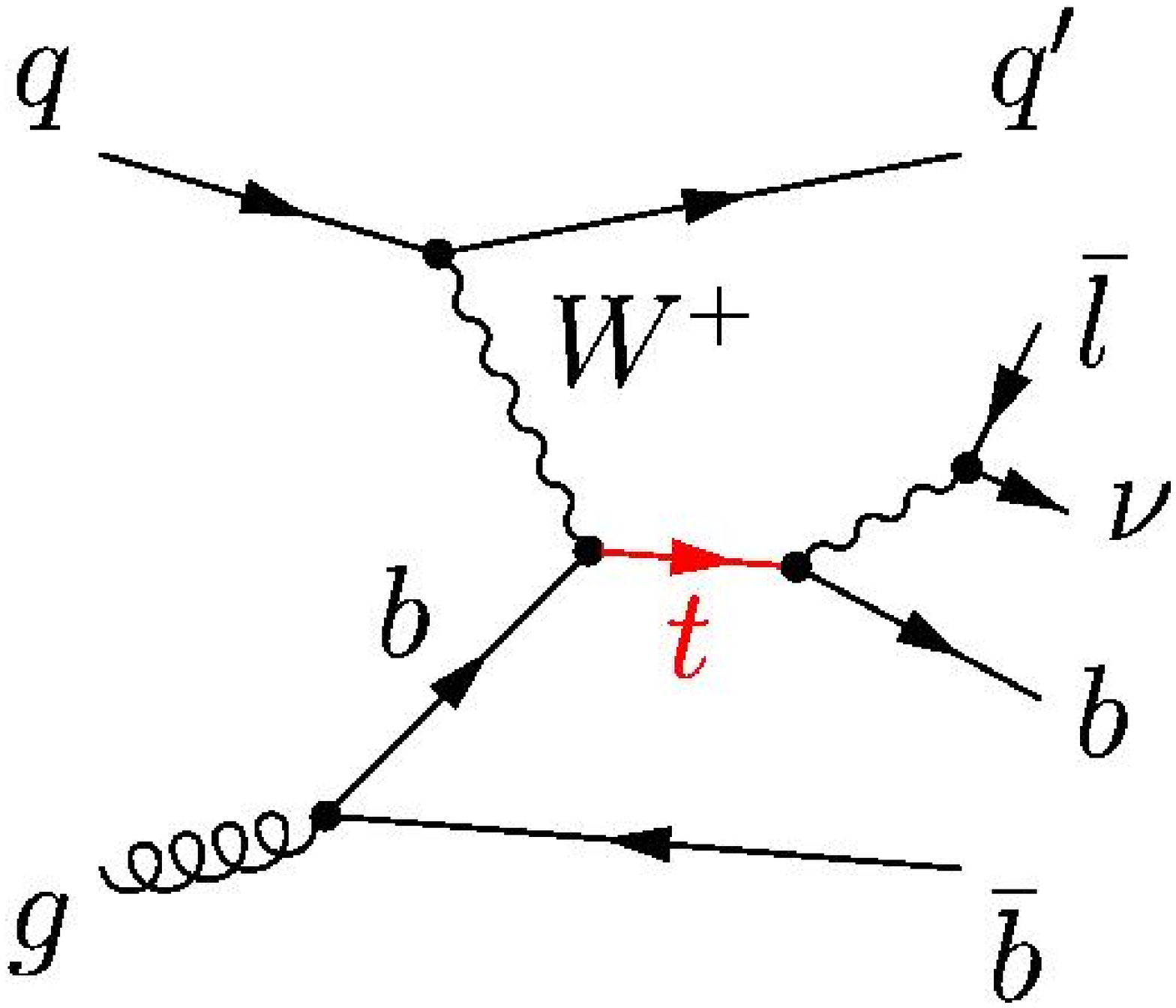}}
  \subfigure[\label{fig:s-channel_NLO} $s$-channel]{
  \includegraphics[width=0.3\textwidth]{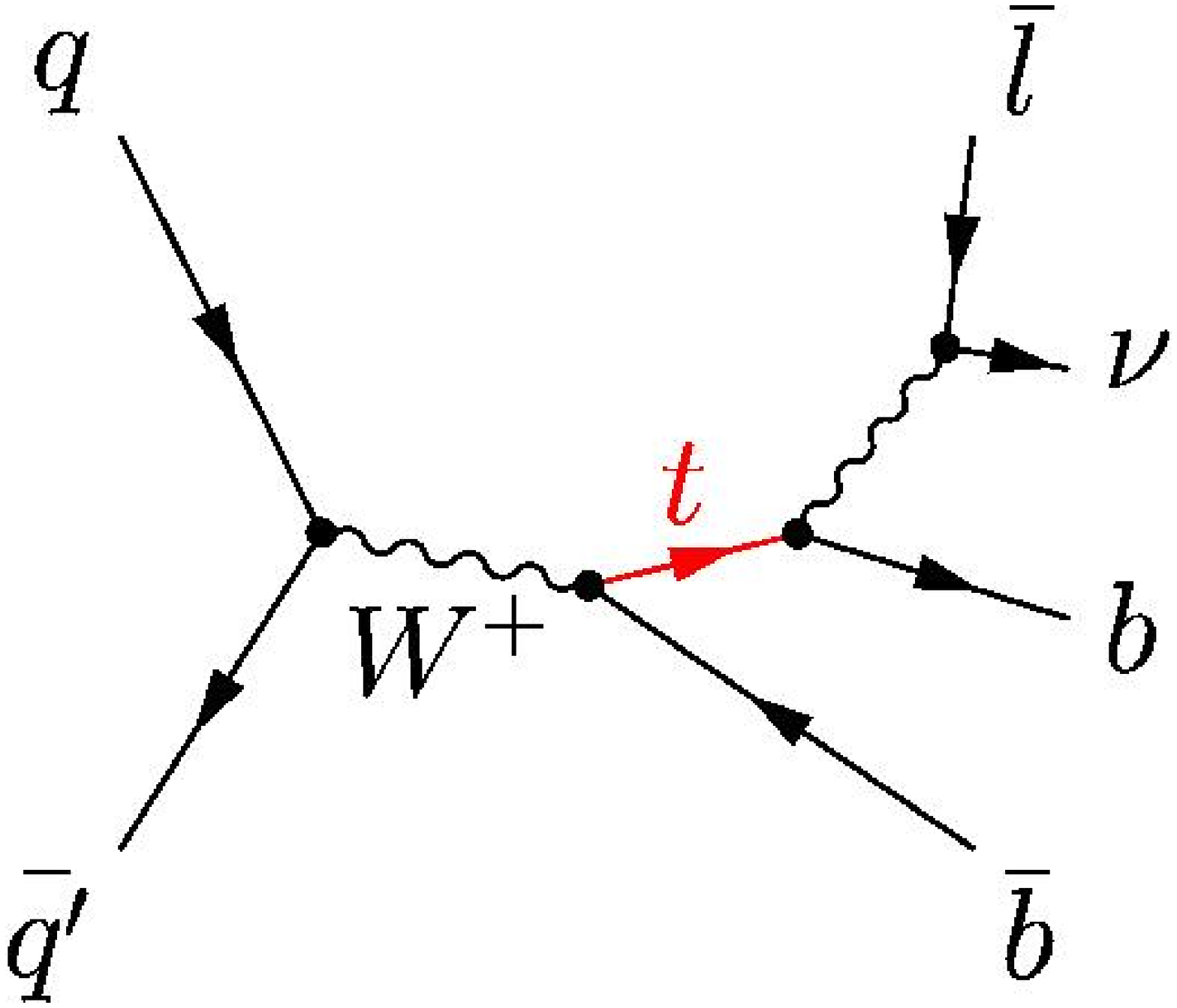}}
  \subfigure[\label{fig:Wt-channel_NLO} $Wt$-channel]{
  \includegraphics[width=0.3\textwidth]{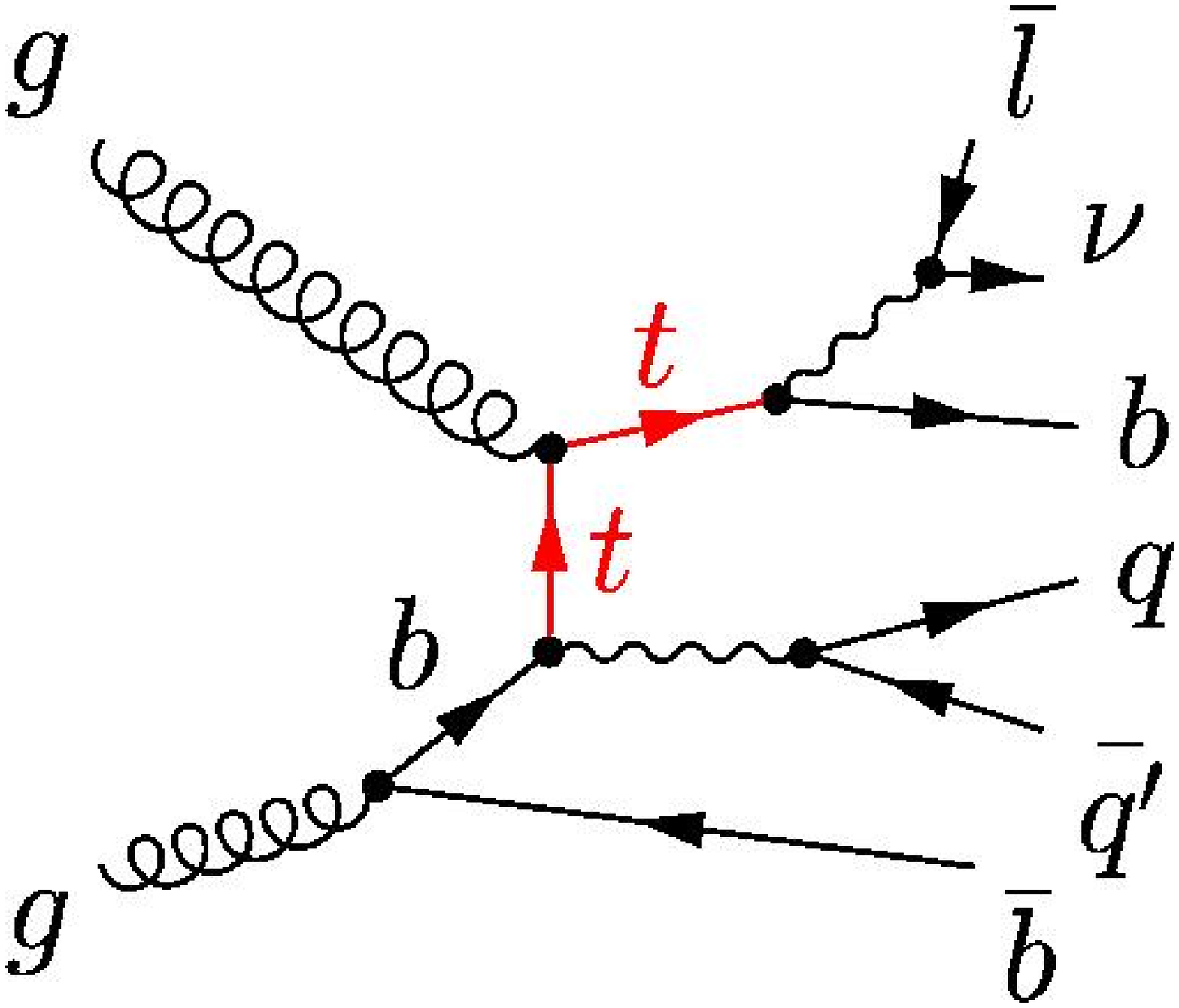}}
  \caption{Representative Feynman diagrams of single top quark production
  channels: (a) $t$-channel $2\rightarrow 3$ process at NLO with initial-state
  gluon splitting, (b) $s$-channel at leading order, and  (c) $Wt$ associated
  production at next-to-leading order (NLO) with initial-state gluon
  splitting.}\label{ST_channels}
\end{figure}

In the Standard Model (SM), the top quark can be produced via the strong
interaction as a $t\bar{t}$ pair. The SM also allows for the top quark to be
produced through the electroweak interaction as a single top quark plus jets.
This is referred to as single top. As illustrated in Figure \ref{ST_channels},
there are three production modes: the $t$-channel ($tbq$), the $s$-channel
($tb$) and the $Wt$-channel processes. Single top quark production was first
observed simultaneously by the CDF and \dzero experiments in
2009.\cite{Aaltonen:2009jj} \cite{PhysRevLett.103.092001} The study of single
top quark events will provide access to the properties of the $Wtb$ coupling.
Within the SM, the single top signal allows for a direct measurement of the
Cabibbo-Kobayashi-Maskawa (CKM) matrix element $V_{tb}$.\cite{Alwall:2006bx}
Furthermore, since the top quark decays before hadronization, its polarization
can be directly observed in the angular correlations of its decay
products.\cite{Mahlon:1996pn} \cite{Mahlon:1999gz}  Single top processes are
expected to be sensitive to several kinds of new physics such as
flavor-changing neutral currents (FCNC).

\section{Event Selection}
The CDF collaboration performed the analysis on a lepton + jets dataset
corresponding to an integrated luminosity of 7.5~fb$^{-1}$. Events are
required to have an isolated electron or muon with $p_{\rm T}$ $>$ 20 GeV,
\MET $>$ 25 GeV, and 2--3 jets with $p_{\rm T}$ $>$ 20 GeV where at least one
jet is $b$-tagged. By selecting a high quality, high-$p_{\rm T}$ isolated
track, the signal acceptance increases by 15\% compared to the previous CDF
analysis.  In this analysis, the CDF collaboration uses the \textsc{powheg}
generator\cite{Alioli:2009je} for single top signal modeling with NLO
accuracy. 

The \dzero collaboration uses 5.4~fb$^{-1}$ of data collected with a logical
OR of many trigger conditions, which together are fully efficient for the
single top quark signal. The main event selection criteria applied is similar
to the previous analysis:\,\cite{PhysRevLett.103.092001} an isolated electron
or muon with $p_{\rm T}$ $>$ 15 GeV, \MET $>$ 20 GeV, 2--4 jets with $p_{\rm
T}$ $>$ 15 GeV out of which one jet has $p_{\rm T}$ $>$ 25 GeV and at least
one jet tagged with a neural-network-based $b$-tagging algorithm. Additional
selection criteria remove multijet background events with misidentified
leptons.

Both collaborations use similar methods for signal and background modeling.
They normalized the $t\bar t$, diboson and $Z$ + jets processes to the SM
prediction.  The QCD models are derived from the data with a non-isolated
lepton (\dzero) or anti-lepton (CDF). Before $b$--tagging, the $W$ + jets and
QCD backgrounds are normalized to the data using the \MET\ variable (CDF) or
several kinematic variables (\dzero). 

\section{Signal-Background Separation}
After the event selection, additional multivariate techniques are used by both
collaborations to further separate signal from backgrounds. Each multivariate
technique constructs a powerful discriminant variable from different input
variables that is proportional to the probability of an event being signal.
The discriminant distribution is used as input to the cross section
measurement. Several validation tests are conducted by studying the
discriminant output distributions in background-enriched control samples.

The CDF collaboration uses a neural network (NN) multivariate technique to
obtain a single top discriminant. By using 11--14 input variables, four
separate NNs are constructed for different analysis channels based on the
number of jets and $b$-tags. For each of the four different channels, the NN
is optimized separately. In the channel with 2 jets and 2 $b$-tags, the NN is
trained for the $s$-channel process as signal without knowledge of the
$t$-channel. The remaining analysis channels are trained for the $t$-channel
process as signal without knowledge of the $s$-channel. To further constrain
the cross section measurement uncertainty, the CDF collaboration trained the
NNs with samples that include a small fraction of events with variations in
the jet energy scale and $Q^{2}$ scales. This method is expected to yield a
3\% improvement in the uncertainty of the single top cross section
measurement.

The \dzero collaboration uses three individual techniques to separate single
top quark events from the background, namely boosted decision trees (BDT),
bayesian neural networks (BNN), and a neuroevolution of augmented topologies
(NEAT).\cite{Abazov:2011pt} All three methods use the same data and background
models, and are trained separately for two channels: for the $tb$
discriminant, which treats the $tb$ process as the signal and the $tqb$
process as a part of the background, and vice versa for the $tbq$
discriminant. With 70\% correlations among the outputs of the individual
methods, a second BNN is used to construct a combined discriminant from the
three discriminant outputs to increase sensitivity. 

\section{Measurement of Cross Section and $|V_{tb}|$}
\begin{figure}[ht!]
\centering
\subfigure[\label{fig:CDF_XS}]{
\includegraphics[width=0.4\textwidth]{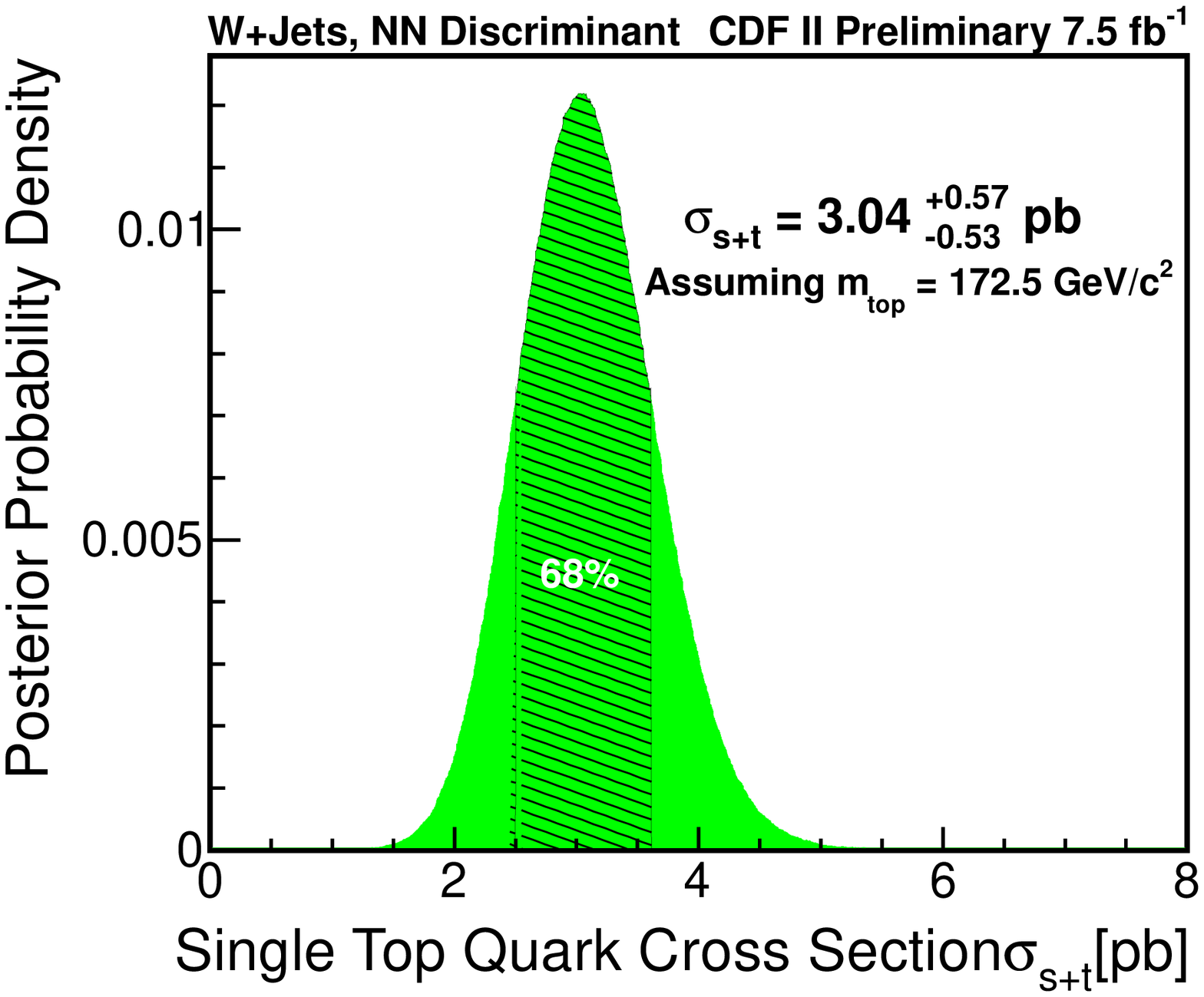}}
\subfigure[\label{fig:D0_XS}]{
\includegraphics[width=0.4\textwidth]{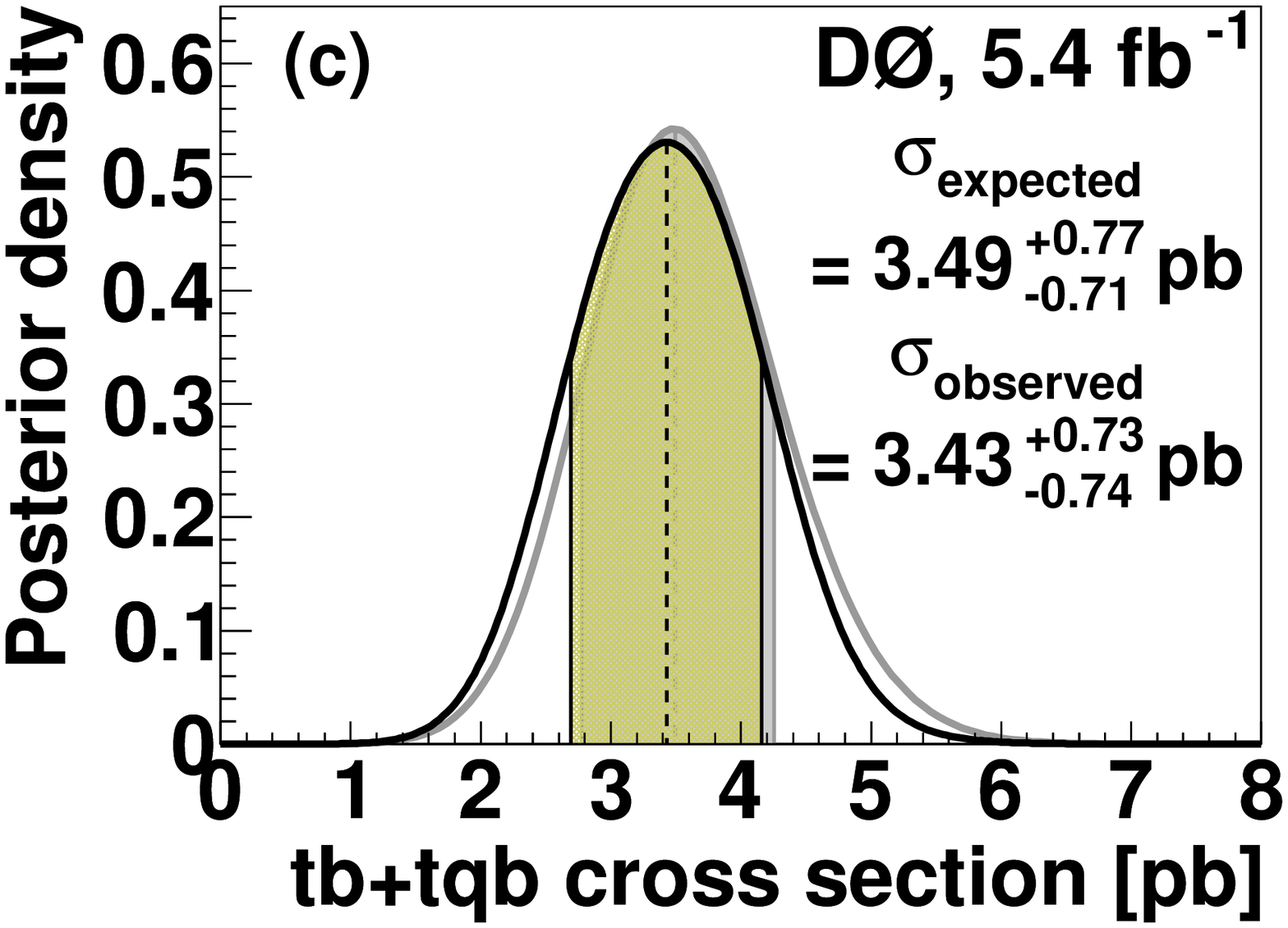}}
\caption{The posterior curve of the cross section measurement for the (a)
CDF and (b) \dzero collaborations.}\label{fig:XS}
\end{figure}

Both experiments measure the single top quark production cross section from
the discriminant output distributions using a Bayesian-binned likelihood
technique. The statistical uncertainty and all systematic uncertainties and
their correlations are considered in these calculations.  The single top cross
section measured by the CDF collaboration is $3.04^{+0.57}_{-0.53}$ pb.  The
\dzero collaboration measured the single top cross section for $tb+tqb$ to be
$3.43^{+0.73}_{-0.74}$ pb.\cite{Abazov:2011pt} Both of the measurements are
shown in Figure \ref{fig:XS}.

Since the single top cross section is directly proportional to $|V_{tb}|^{2}$,
both collaborations use the cross section measurements to extract $|V_{tb}|$.
By restricting the measurement to the SM interval [0, 1], CDF measures
$|V_{tb}|$ = 0.96 $\pm$ 0.09 (stat+syst) $\pm$ 0.05 (theory), and sets a limit
of $|V_{tb}|$  $>$ 0.78 at 95\% C.L. Using the same interval [0, 1], \dzero
extracts the limit of $|V_{tb}|$ $>$ 0.79 at 95\% C.L.; after removing the
upper constraint of the interval, \dzero measures $|V_{tb}f^{L}_{1}|$ =
$1.02^{+0.10}_{-0.11}$, where $f^{L}_{1}$ is the strength of the left-handed
$Wtb$ coupling.

\section{$t$-Channel Observation}
With the same multivariate discriminant for the $t$-channel process, \dzero
computes the significance of the $t$-channel cross section using a
log-likelihood ratio approach, which tests the compatibility of the data with
two hypotheses: a null hypothesis with only background and a test hypothesis
with background plus signal. The computation of the distributions for these
two hypotheses is given by an asymptotic Gaussian approximation. With this
approximation, the significance of the measured $t$-channel cross section is
independent of any assumption on the production rate of
$s$-channel.\cite{Abazov:2011rz} The estimated probability corresponds to an
observed significance of 5.5 standard deviations with an expected significance
of 4.6 standard deviations.

\section{\label{sec:2D} Two-Dimensional Fit Results} 
\begin{figure}[ht!]
\centering
\subfigure[\label{fig:CDF_2D}]{
\includegraphics[width=0.35\textwidth]{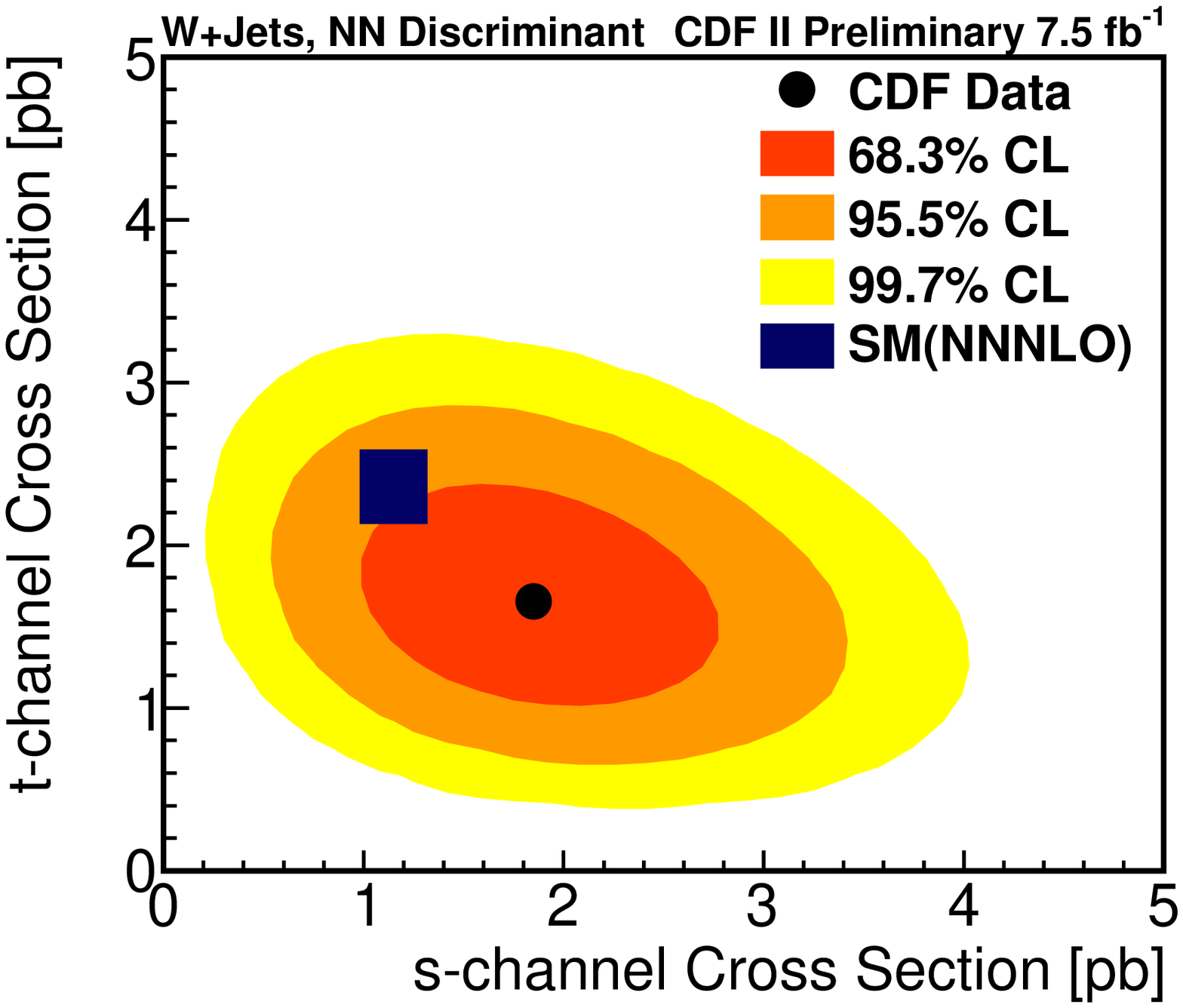}}
\subfigure[\label{fig:D0_2D}]{
\includegraphics[width=0.35\textwidth]{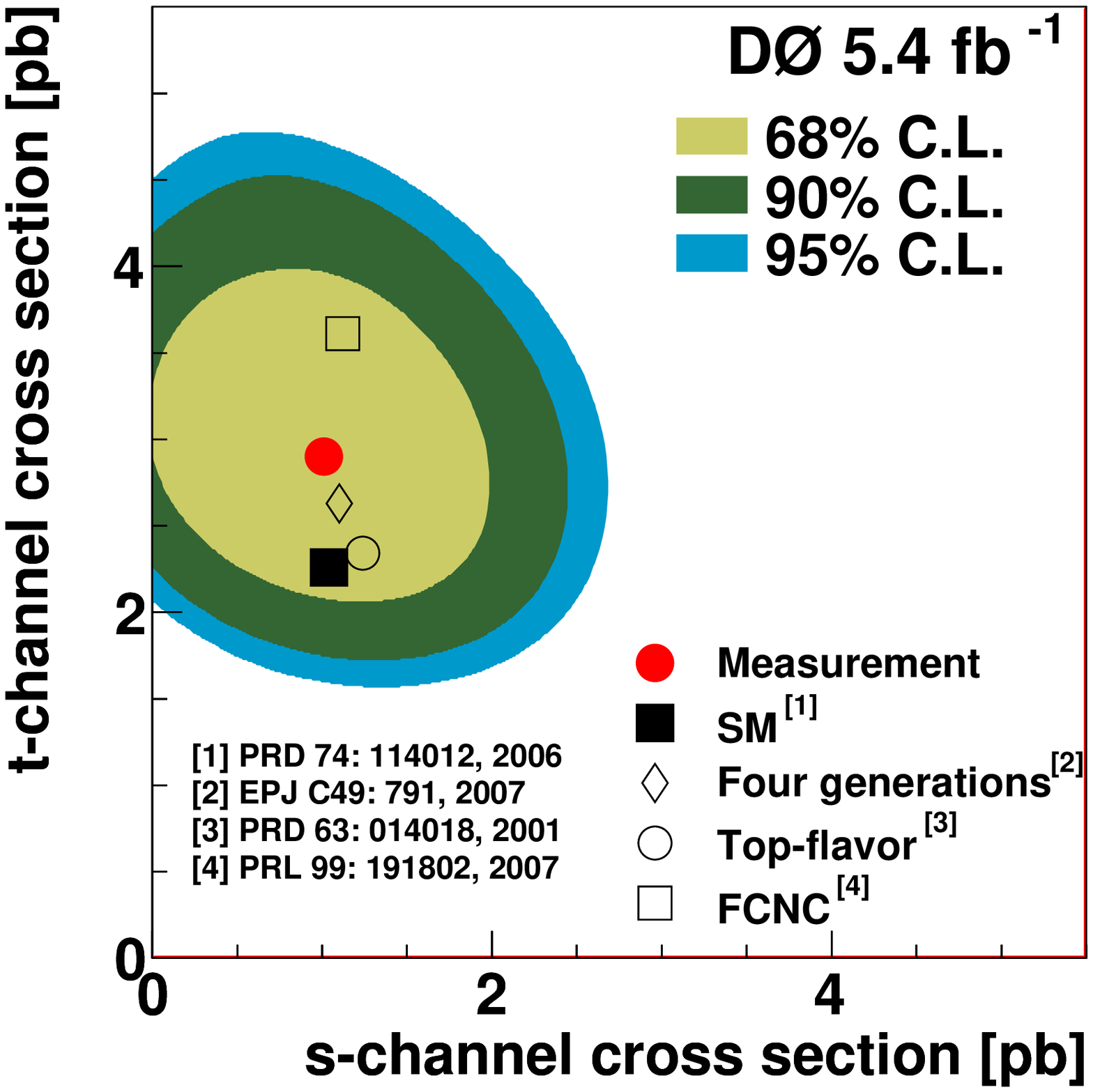}}
\caption{ The results of the two-dimensional fit for 
    $\sigma_{s}$ and $\sigma_{t}$.  The circle point
    shows the best fit value and the 68\%, 95\%, and 99\% credibility
    regions are shown as shaded areas. The SM prediction is also
    indicated with theoretical uncertainties included. }\label{fig:2D}
\end{figure}

The combined signal cross section ($\sigma_{s+t}$) is extracted by
constructing a one-dimensional Bayesian posterior. An extension is to form a
two-dimensional posterior probability density as a function of the cross
sections for the $s$- and $t$-channel as in Figure \ref{fig:2D}. The best-fit
cross section is the one for which the posterior is maximized without assuming
the SM-predicted ratio between the cross section for the $s$- and
$t$-channels. 

\section{Anomalous $Wtb$ Coupling}
Within the SM theory, the top quark coupling to the bottom quark and the $W$
boson ($tWb$) has the V--A form of a left-handed vector interaction.
Deviations from the SM expectation in the coupling form factors can manifest
themselves by altering the fraction of $W$ boson from top quark decays or by
changing the rate and kinematic distributions of electroweak single top quark
production. Three separate scenarios are investigated and upper limits are set
with the same dataset by \dzero for $f^{R}_{V}$, $f^{L}_{T}$, and
$f^{R}_{T}$.\cite{Abazov:2011pm}

\section{Summary}
The CDF and \dzero collaborations have performed precise measurements of the
electroweak single top quark production cross section and the CKM matrix
element $|V_{tb}|$ using 7.5 and 5.4~fb$^{-1}$ of data, respectively. An
anomalous $Wtb$ coupling search by \dzero investigates three separate
scenarios and sets an upper limit on each of them. 

\section*{Acknowledgments}
We thank the Fermilab staff, the technical staffs of the participating
institutions, and the funding agencies for their vital contributions.

\end{document}